\documentclass[english,showpacs,floatfix,preprint,10pt]{revtex4}
\usepackage[dvips]{graphicx}
\usepackage{longtable}
\usepackage{amsmath,amssymb}
\usepackage{dcolumn}
\usepackage{latexsym}
\usepackage{babel}
\usepackage[latin1]{inputenc}
\usepackage{color}
\usepackage[colorlinks]{hyperref}

\begin{document}
\baselineskip=0.8 cm
\title{{\bf The imprint of the interaction between dark sectors in large scale cosmic microwave background anisotropies}}

\author{Jian-Hua He$^{1}$, Bin Wang$^{1}%
$\footnote{wangb@fudan.edu.cn }, Pengjie Zhang$^{2}$}
\affiliation{$^{1}$ Department of Physics, Fudan University, 200433
Shanghai, China} \affiliation{$^{2}$ Key Laboratory for Research in Galaxies and Cosmology, Shanghai
  Astronomical Observatory, Nandan Road 80, Shanghai, 200030, China}

\vspace*{0.2cm}
\begin{abstract}
\baselineskip=0.6 cm

Dark energy interacting with dark matter is a promising model to
solve the cosmic coincidence problem. We study the signature of such
interaction on large scale cosmic microwave background (CMB)
temperature anisotropies. Based on the detail analysis in
perturbation equations of dark energy and dark matter when they are
in interaction, we find that the large scale CMB, especially the
late Integrated Sachs Wolfe effect, is a useful tool to measure the
coupling between dark sectors. We also discuss the possibility to
detect the coupling by cross-correlating CMB maps with tracers of the large
scale structure.  We finally perform the global fitting to
constrain the coupling by using the CMB power spectrum data together with other
observational data. We find that in the $1\sigma$ range, the
constrained coupling between dark sectors can solve the coincidence
problem.

\end{abstract}

 \maketitle
\newpage
\section{Introduction}

Advances in cosmological observations have presented us a
concordance picture that our present universe is accelerating and
the dominant mass-energy components for the evolution of the
universe in the standard model are non-baryonic cold dark matter
(DM) and dark energy field (DE)\cite{1,2,3}. The DE is identified as
the engine for the accelerated expansion. The leading interpretation
of such DE is a cosmological constant with equation of state (EoS)
$w =-1$. While the cosmological constant is consistent with the
observational data, at the fundamental level it fails to be
convincing: the vacuum energy density falls  below the value
predicted by any sensible quantum field theory by many orders of
magnitude, and it unavoidably leads to the coincidence problem,
i.e., ``why are the vacuum and matter energy densities of precisely
the same order today?". More sophisticated models have been proposed
to replace the cosmological constant by a dynamical dark energy in
the conjectures relating the DE either to a scalar field called
quintessence with $w >-1$, or to an exotic field called phantom with
$w <-1$. However, there is no clear winner in sight to explain the
nature of DE at the moment.

Most available discussions on the DE were in the minimal picture
which assumed that DE does not feel any significant interactions
from DM. Although this picture is consistent with current
observations, considering that DE and DM account for significant
fractions of the content of the Universe, it is still natural, in
the framework of field theory, to consider their interaction. The
possibility that DE and DM can interact with each other has been
widely discussed recently\cite{10}-\cite{31}. It has been shown that
the coupling between DE and DM can provide a mechanism to alleviate
the coincidence problem \cite{10, 11, 13, 14}.  Furthermore, it was
suggested that the dynamical equilibrium of collapsed structures
such as clusters would get modification due to the coupling between
DE and DM \cite{19, 24}. The influences on the growth of cosmic
structure due to the coupling between dark sectors were also shown
in \cite{31,z}. Observational signatures on the dark sectors' mutual
interaction have been found in the probes of the cosmic expansion
history by using the SNIa, BAO and CMB shift data etc \cite{20, 25,
26}. In addition, it has been argued that an appropriate interaction
between DE and DM can influence the perturbation dynamics and affect
the lowest multipoles of the CMB power spectrum \cite{15, 16}.
Recently there has been some concern about the stability of the
perturbations if DE and DM interact \cite{21}. However, it was
proved in \cite{22} that the stability of the curvature perturbation
depends on the types of coupling between dark sectors and the EoS of
DE. Based on the result in \cite{22}, we are in a position to carry
out the global fitting to probe the interaction between dark sectors
by using the CMB power spectrum data together with other
observational data.

Recently the WMAP data showed the deficit of large scale power in
the temperature map, in particular in the CMB quadrupole. The
significant contribution to the fluctuations on these scales is the
late Integrated Sachs Wolfe (ISW) effect which is induced by the
passage of CMB photons through the time evolving gravitational
potential when the universe enters a rapid expansion phase once DE
dominates. The late ISW effect has the unique ability to probe the
``size" of DE. Much effort has been put into determining the EoS and
the speed of sound of DE \cite{50,51}. Whether the late ISW can give
insight into the coupling between dark sectors is an interesting
question. In this paper we are going to discuss this problem in
detail. We will further employ the CMB data from ground based and
satellite observations together with SNIa and SDSS data to constrain
the coupling between dark sectors.

The organization of the paper is as follows: in the following
section we will review the general formalism of the perturbation
theory in the presence of the interaction between DE and DM. In
Sec.III we will discuss the large scale cosmic microwave
anisotropies and its imprint on the ``size" of DE, especially the
coupling between dark sectors. In Sec.IV, we will present the global
fitting result by using CMB data together with SNIa and
SDSS data and we will discuss the alleviation of the coincidence
problem when DE interacts with DM. We will present our conclusions
and discussions in the end.

\section{perturbation theory when DE interacts with DM}

In this section we will go over the perturbation theory when DE
interacts with DM. The detailed descriptions can be found in
\cite{22,31}.

The perturbed metric at first order is of the form
\begin{equation}
ds^2 = a^2[-(1+2\psi)d\tau^2+2\partial_iBd\tau
dx^i+(1+2\phi)\delta_{ij}dx^idx^j+D_{ij}Edx^idx^j],\label{perturbedspacetime}
\end{equation}
where $\psi, B, \phi, E$ represent the scalar metric perturbations,
$a$ is the cosmic scale factor and
\begin{equation}
D_{ij}=(\partial_i\partial_j-\frac{1}{3}\delta_{ij}\nabla^2).
\end{equation}

We work with general stress-energy tensor
\begin{equation}
T^{\mu\nu}=\rho U^{\mu}U^{\nu}+p(g^{\mu\nu}+U^{\mu}U^{\nu}),
\end{equation}
for a two-component system consisting of DE and DM. Each
energy-mementum tensor satisfies the conservation law
\begin{eqnarray}
\nabla_{\mu}T^{\mu\nu}_{(\lambda)}&=Q^{\nu}_{(\lambda)},
\end{eqnarray}
where $Q^{\nu}_{(\lambda)}$ denotes the interaction between
different components and $\lambda$ denotes either the DM or the DE
sector.

In the Fourier space the perturbed energy-momentum tensor reads
\cite{31}
\begin{eqnarray}
\delta_{\lambda}'+3\mathcal{H}(\frac{\delta p_{\lambda}}{\delta
\rho_{\lambda}}-w_{\lambda})\delta_{\lambda}=
-(1+w_{\lambda})kv_{\lambda}
-3(1+w_{\lambda})\phi'+(2\psi-\delta_{\lambda})\frac{a^2Q^0_{\lambda}}{\rho_{\lambda}}+\frac{a^2
\delta Q^0_{\lambda}}{\rho_{\lambda}}\nonumber\\\label{perturbed}
(v_{\lambda}+B)'+\mathcal{H}(1-3w_{\lambda})(v_{\lambda}+B) =
\frac{k}{1+w_{\lambda}}\frac{\delta p_{\lambda}}{\delta
\rho_{\lambda}}\delta_{\lambda}
-\frac{w_{\lambda}'}{1+w_{\lambda}}(v_{\lambda}+B)+k\psi
-\frac{a^2Q^0_{\lambda}}{\rho_{\lambda}}v_{\lambda}-\frac{w_{\lambda}a^2Q^0_{\lambda}}{(1+w_{\lambda})\rho_{\lambda}}B+\frac{a^2\delta
Q_{p\lambda}}{(1+w_{\lambda})\rho_{\lambda}}.
\end{eqnarray}

Constructing the gauge invariant quantities\cite{31}
\begin{eqnarray}
\Psi&=& \psi -
\frac{1}{k}\mathcal{H}(B+\frac{E'}{2k})-\frac{1}{k}(B'+\frac{E^{''}}{2k})\nonumber\\
\Phi&=&\phi+\frac{1}{6}E-\frac{1}{k}\mathcal{H}(B+\frac{E'}{2k})\nonumber\\
Q^{0I}_{\lambda }&=&\delta
Q^0_{\lambda}-\frac{Q^{0'}_{\lambda}}{\mathcal{H}}(\phi+\frac{E}{6})+Q^{0}_{\lambda}\left[\frac{1}{\mathcal{H}}(\phi+\frac{E}{6})\right]'\label{invariantQ}\\
\delta Q_{p\lambda}^{I}&=&\delta Q_{p\lambda }-Q^0_{\lambda
}\frac{E'}{2k}\nonumber\\
 D_{g\lambda}&=&
\delta_{\lambda}-\frac{\rho_{\lambda}'}{\rho_{\lambda}\mathcal{H}}\left(\phi+\frac{E}{6}\right)\nonumber\\
V_{\lambda} &=& v_{\lambda} -\frac{E'}{2k},\nonumber
\end{eqnarray}
we obtain the gauge invariant linear perturbation equations for dark
sectors. For DM it reads,
\begin{eqnarray}
&&D'_{gc}+\left \{
\left(\frac{a^2Q_c^0}{\rho_c\mathcal{H}}\right)'+\frac{\rho_c'}{\rho_c\mathcal{H}}\frac{a^2Q_c^0}{\rho_c}
\right \}\Phi + \frac{a^2 Q_c^0}{\rho_c}D_{gc}+
\frac{a^2Q_c^0}{\rho_c\mathcal{H}}\Phi' \nonumber \\
 &&= -kV_c +2\Psi
\frac{a^2Q_c^0}{\rho_c}+\frac{a^2\delta
Q_c^{0I}}{\rho_c}+\frac{a^2Q_c^{0'}}{\rho_c\mathcal{H}}\Phi-\frac{a^2Q_c^0}{\rho_c}\left(\frac{\Phi}{\mathcal{H}}\right)'\nonumber\label{DMV}
\\
&&V_c'+\mathcal{H}V_c=k\Psi-\frac{a^2Q_c^0}{\rho_c}V_c+\frac{a^2\delta
Q_{pc}^{I}}{\rho_c}
\end{eqnarray}
For DE, we have
\begin{eqnarray}
&&D'_{gd}+\left\{\left(\frac{a^2Q_d^0}{\rho_d\mathcal{H}}\right)'-3w'+3(C_e^2-w)\frac{\rho_d'}{\rho_d}+\frac{\rho_d'}{\rho_d\mathcal{H}}\frac{a^2Q_d^0}{\rho_d}\right\}\Phi+\left\{3\mathcal{H}(C_e^2-w)+\frac{a^2Q_d^0}{\rho_d}\right\}D_{gd}+\frac{a^2Q_d^0}{\rho_d\mathcal{H}}\Phi'\nonumber\\
&=&-(1+w)kV_d+3\mathcal{H}(C_e^2-C_a^2)\frac{\rho_d'}{\rho_d}\frac{V_d}{k}+2\Psi\frac{a^2Q_d^0}{\rho_d}+\frac{a^2\delta
Q_d^{0I}}{\rho_d}+\frac{a^2Q_d^{0'}}{\rho_d\mathcal{H}}\Phi-\frac{a^2Q_d^0}{\rho_d}\left(\frac{\Phi}{\mathcal{H}}\right)'\nonumber\label{DEV}\\
&&V_d'+\mathcal{H}(1-3w)V_d=\frac{kC_e^2}{1+w}D_{gd}+\frac{kC_e^2}{1+w}\frac{\rho_d'}{\rho_d\mathcal{H}}\Phi-\left(C_e^2-C_a^2\right)\frac{V_d}{1+w}\frac{\rho_d'}{\rho_d}-\frac{w'}{1+w}V_d+k\Psi-\frac{a^2Q_d^0}{\rho_d}V_d+\frac{a^2\delta
Q_{pd}^{I}}{\rho_d},
\end{eqnarray}
where we have employed
\begin{equation}
\frac{\delta p_d}{\rho_d} =
C_e^2\delta_d-(C_e^2-C_a^2)\frac{\rho_d'}{\rho_d}\frac{v_d+B}{k}
\end{equation}
where $C_e^2$ is the effective sound speed of DE at the rest frame,
$C_a^2$ is the adiabatic sound speed.

To alleviate the singular behavior caused by $w$ crossing $-1$, we
substitute $V_{\lambda}$ into $U_{\lambda}$ in the above equations
where
\begin{equation}
U_{\lambda}=(1+w)V_{\lambda}.
\end{equation}
Thus we can rewrite eq(~\ref{DMV}) and eq(~\ref{DEV}) into,
\begin{eqnarray}
&&D'_{gc}+\left \{
\left(\frac{a^2Q_c^0}{\rho_c\mathcal{H}}\right)'+\frac{\rho_c'}{\rho_c\mathcal{H}}\frac{a^2Q_c^0}{\rho_c}
\right \}\Phi + \frac{a^2 Q_c^0}{\rho_c}D_{gc}+
\frac{a^2Q_c^0}{\rho_c\mathcal{H}}\Phi' \nonumber \label{DMU} \\
 &&= -kU_c +2\Psi
\frac{a^2Q_c^0}{\rho_c}+\frac{a^2\delta
Q_c^{0I}}{\rho_c}+\frac{a^2Q_c^{0'}}{\rho_c\mathcal{H}}\Phi-\frac{a^2Q_c^0}{\rho_c}\left(\frac{\Phi}{\mathcal{H}}\right)'\nonumber
\\
&&U_c'+\mathcal{H}U_c=k\Psi-\frac{a^2Q_c^0}{\rho_c}U_c+\frac{a^2\delta
Q_{pc}^{I}}{\rho_c}
\end{eqnarray}
\begin{eqnarray}
&&D'_{gd}+\left\{\left(\frac{a^2Q_d^0}{\rho_d\mathcal{H}}\right)'-3w'+3(C_e^2-w)\frac{\rho_d'}{\rho_d}+\frac{\rho_d'}{\rho_d\mathcal{H}}\frac{a^2Q_d^0}{\rho_d}\right\}\Phi+\left\{3\mathcal{H}(C_e^2-w)+\frac{a^2Q_d^0}{\rho_d}\right\}D_{gd}+\frac{a^2Q_d^0}{\rho_d\mathcal{H}}\Phi'\nonumber\\
&=&-kU_d+3\mathcal{H}(C_e^2-C_a^2)\frac{\rho_d'}{\rho_d}\frac{U_d}{(1+w)k}+2\Psi\frac{a^2Q_d^0}{\rho_d}+\frac{a^2\delta
Q_d^{0I}}{\rho_d}+\frac{a^2Q_d^{0'}}{\rho_d\mathcal{H}}\Phi-\frac{a^2Q_d^0}{\rho_d}\left(\frac{\Phi}{\mathcal{H}}\right)'\nonumber \label{DEU}\\
&&U_d'+\mathcal{H}(1-3w)U_d=kC_e^2D_{gd}+kC_e^2\frac{\rho_d'}{\rho_d\mathcal{H}}\Phi-\left(C_e^2-C_a^2\right)\frac{U_d}{1+w}\frac{\rho_d'}{\rho_d}+(1+w)k\Psi-\frac{a^2Q_d^0}{\rho_d}U_d+(1+w)\frac{a^2\delta
Q_{pd}^{I}}{\rho_d}.
\end{eqnarray}
The quantity $\Phi$ is given by,
\begin{equation}
\Phi=\frac{4\pi Ga^2\sum \rho_i\{D_g^i+3\mathcal{H}U^i/k\}}{k^2-4\pi
Ga^2\sum \rho_i'/\mathcal{H}}\quad.
\end{equation}

To solve the above equations we have to specify the interaction form
$Q^{\nu}$ between DE and DM. However, this is a hard task, since the
nature of DE and DM remains unknown, it is not possible at the
present moment to derive the precise form of the interaction between
them from first principles. One has to assume a specific coupling
from the outset \cite{11, 29, 30} or determine it from
phenomenological requirements \cite{12, 25}. For the generality, we
can assume the phenomenological description of the interaction
between dark sectors in the comoving frame\cite{22,31}
\begin{eqnarray}
Q_c^{\nu}=\left[\frac{3\mathcal{H}}{a^2}(\xi_1\rho_c+\xi_2\rho_d),0,0,0\right]^{T}\nonumber\\
Q_d^{\nu}=\left[-\frac{3\mathcal{H}}{a^2}(\xi_1\rho_c+\xi_2\rho_d),0,0,0\right]^{T},
\end{eqnarray}
where $\xi_1, \xi_2$ are small positive dimensionless constants and
$T$ is the transpose of the matrix. Choosing positive sign in the
interaction, one can ensure the direction of energy transfer from DE
to DM, which is required to alleviate the coincidence problem
\cite{17, 23} and avoid some unphysical problems such as negative DE
density etc \cite{21, 25}. The perturbed gauge-invariant coupling
vector can be calculated by
\begin{equation}
\frac{a^2\delta
Q_c^{0I}}{\rho_c}=3\mathcal{H}\{\xi_1D_{gc}+\xi_2D_{gd}/r\}-3\left\{(\frac{\mathcal{H}'}{\mathcal{H}}-2\mathcal{H})(\xi_1+\xi_2/r)\right\}\Phi+\frac{a^2Q_c^0}{\rho_c}\left[
\frac{\Phi}{\mathcal{H}}\right]'
\end{equation}
\begin{equation}
\frac{a^2\delta
Q_d^{0I}}{\rho_d}=-3\mathcal{H}\{\xi_1D_{gc}r+\xi_2D_{gd}\}+3\left\{(\frac{\mathcal{H}'}{\mathcal{H}}-2\mathcal{H})(\xi_1r+\xi_2)\right\}\Phi+\frac{a^2Q_d^0}{\rho_d}\left[
\frac{\Phi}{\mathcal{H}}\right]'
\end{equation}
where $r=\rho_c/\rho_d$. For the reason as illustrated in
\cite{22,31}, we set $\delta Q_{pm}^{I}=\delta Q_{pd}^{I}=0$. Using
continuous equations to eliminate $\rho_{\lambda}'/\rho_{\lambda}$
in eq(~\ref{DMU}), eq(~\ref{DEU}), we obtain the perturbation
equations\cite{22},
\begin{eqnarray}
D_{gc}'&=&-kU_c+6\mathcal{H}\Psi(\xi_1+\xi_2/r)-3(\xi_1+
\xi_2/r)\Phi'+3\mathcal{H}\xi_2(D_{gd}-D_{gc})/r\quad ,\\
U_c'&=&-\mathcal{H}U_c+k\Psi-3\mathcal{H}(\xi_1+\xi_2/r)U_c\quad ,\\
D_{gd}' & = & -3\mathcal{H}C_e^2\left\{D_{gd}-3\left(\xi_1r+\xi_2+
1+w\right )\Phi\right\}-3\mathcal{H}(C_e^2-C_a^2)
\left[\frac{3\mathcal{H}U_d}{k}-a^2Q^0_d\frac{U_d}{(1+w)\rho_dk}\right]
 \nonumber\\
&&-9\mathcal{H}w\left(\xi_1r+\xi_2+1+w\right) \Phi+3\mathcal{H}w
D_d+3w'\Phi+3(\xi_1r+\xi_2)\Phi'-kU_d-6\Psi\mathcal{H}(\xi_1
r+\xi_2)\nonumber\\
&&+3\mathcal{H}\xi_1r(D_{gd}-D_{gc}) \label{density} \\
U_d'& = &-\mathcal{H}(1-3w)U_d+kC_e^2\left\{D_{gd}-3\left(\xi_1
r+\xi_2+1+w\right )\Phi\right\} \nonumber\\
&&-(C_e^2-C_a^2)a^2Q^0_d\frac{U_d}{(1+w)\rho_d}+
3(C_e^2-C_a^2)\mathcal{H}U_d+(1+w)k\Psi+3\mathcal{H}
(\xi_1r+\xi_2)U_d. \label{velocity}
\end{eqnarray}
In solving these equations we will adopt the adiabatic initial
condition\cite{22}
\begin{equation}
\frac{D_{gc}}{1-\xi_1-\xi_2/r}=\frac{D_{gd}}{1+w+\xi_1r+\xi_2}.
\end{equation}

Dynamical stability of the DE and DM perturbations were examined in
detail in \cite{22,31}. In the general form of the phenomenological
interaction, when DE EoS $w>-1$ and we choose the coupling between
dark sectors proportional to the DM energy density (by setting
$\xi_2 = 0$ while keeping $\xi_1$ nonzero) or the total dark sectors
energy density (by setting $\xi_1=\xi_2$), we observed the dynamical
instability in perturbations as argued in \cite{21}. However, when
we choose the dark sectors' mutual interaction proportional to the
energy density of DE by taking $\xi_1 = 0$ and $\xi_2 \neq 0$, even
for $w>-1$ we have the stable perturbations. When EoS of DE $w<-1$,
no matter the interaction proportional to the energy density of the
individual dark sector or the total dark sectors, the perturbation
is always stable.

\section{large scale cosmic microwave}

With the formalism of the perturbation theory, we are in a position
to study the CMB power spectrum. In our analysis we will concentrate on
models with constant $w$ and a constant speed of sound. The temperature
anisotropy power spectrum can be calculated by
\begin{equation}
C_l=4\pi \int \frac{dk}{k}\mathcal{P}_{\chi}(k)\mid \Delta_l(k,
\tau_0) \mid^2,
\end{equation}
where $\Delta_l$ gives the transfer function for each $l$,
$\mathcal{P}_{\chi}$ is the primodial power spectrum and $\tau_0$ is
the conformal time today. On large scales the transfer functions are
of the form
\begin{equation}
\Delta_l(k,\tau_0)=\Delta_l^{SW}(k)+\Delta_l^{ISW}(k),
\end{equation}
where $\Delta_l^{SW}(k)$ is the contribution from the last
scattering surface given by the ordinary Sachs-Wolfe (SW) effect and
$\Delta_l^{ISW}(k)$ is the contribution due to the change of the
gravitational potential when photons passing through the universe on
their way to earth and is called the ISW effect. The ISW
contribution can be written as
\begin{equation}
\Delta_l^{ISW}= \int_{\tau_i}^{\tau_0} d\tau
j_l(k[\tau_0-\tau])e^{\kappa(\tau_0)-\kappa(\tau)}[\Psi'-\Phi'],
\end{equation}
where $j_l$ is the spherical Bessel function, $\kappa$ denotes the
optical depth for Thompson scattering. From Einstein equations, we
obtain,
\begin{eqnarray}
\Psi' - \Phi' &=& 2\mathcal{H}\left[ \Phi + 4\pi Ga^2 \sum_{i}
U^{i}\rho^i/(\mathcal{H}k) + \mathcal{T} \right ]- \mathcal{T}'
\end{eqnarray}
where
\begin{eqnarray}
\Phi' &=& -\mathcal{H}\Phi - \mathcal{H}\mathcal{T} - 4\pi Ga^2
\sum_{i} U^{i}\rho^i/k \nonumber\\
\mathcal{T}& = &\frac{8 \pi G a^2}{k^2} \left \{
p^{\gamma}\Pi^{\gamma}+ p^{\nu} \Pi ^{\nu}\right \}\nonumber
\end{eqnarray}
and $\Pi$  is the anisotropic stress of relativistic components
which can be neglected in the following discussion.

The ISW effect can be classified into early and late times effects.
The early ISW effect takes place from the time following
recombination to the time when radiation is no longer dynamically significant,
which gives clues about what is happening in the universe from the
radiation domination to matter domination. The late time ISW
effect arises when DE becomes dynamically important and has the unique
ability to probe the ``size" of DE. When DE becomes non-negligible, the
gravitational potential decays. When a photon passes throw a decaying
potential well, it will have a net gain in energy and thus leads to the late
ISW effect. The late ISW effect is a significant contribution to the
large scale power in the temperature map of CMB.

In the minimal case  when there is no interaction between DE and DM,
the presence of DE perturbations leaves a $w$ and $C_e^2$ dependent
signature in the late ISW source term. For $w>-1$, it was observed
that the CMB temperature anisotropies on large scales got enhanced
for bigger constant values of $C_e^2$\cite{50,51}. It was argued
that increasing $C_e^2$ increases the suppression of DM
perturbations and therefore increases the contribution to the ISW
effect\cite{51}. However, when $w<-1$ the effect is reversed, with
the perturbation initially of opposite sign, and the contribution to
the ISW effect increases as the sound speed of DE is
decreased\cite{50}. The interplay between perturbations in the DE
and DM and the ISW effect is very subtle and it is more direct to
cross-correlate the late ISW effect to its source term, the change
of the gravitational potential\cite{50}.

The late ISW effect is a promising tool to measure the EoS and sound
speed of DE. Whether it can present us the signature of the
interaction between DE and DM is a question we are going to address
here. We will concentrate our discussions on three commonly studied
phenomenologically interaction forms between dark sectors, namely
the interaction proportional to the energy density of DE, DM and
total dark sectors, respectively.

\begin{figure}
\begin{center}
  \begin{tabular}{cc}
\includegraphics[width=3.2in,height=3.2in]{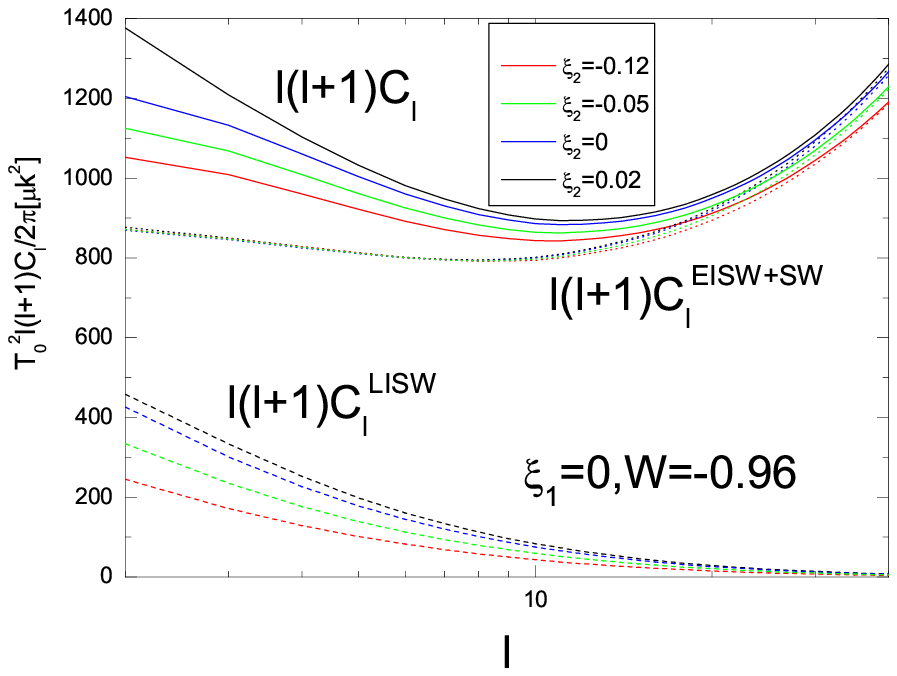}&
\includegraphics[width=3.2in,height=3.2in]{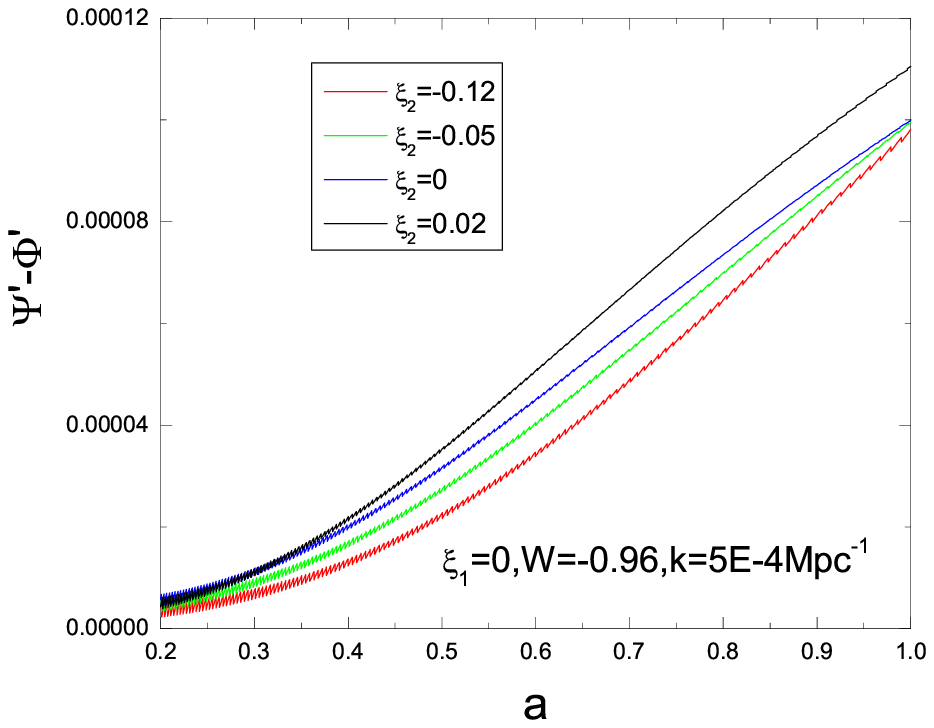}\nonumber \\
    (a)&(b)\nonumber \\
\includegraphics[width=3.2in,height=3.2in]{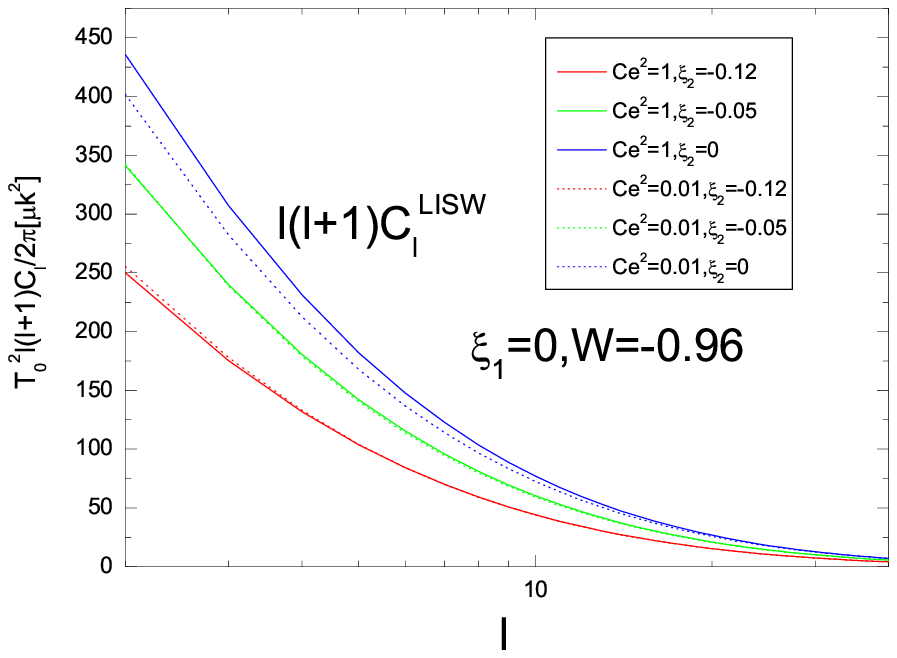}&\includegraphics[width=3.2in,height=3.2in]{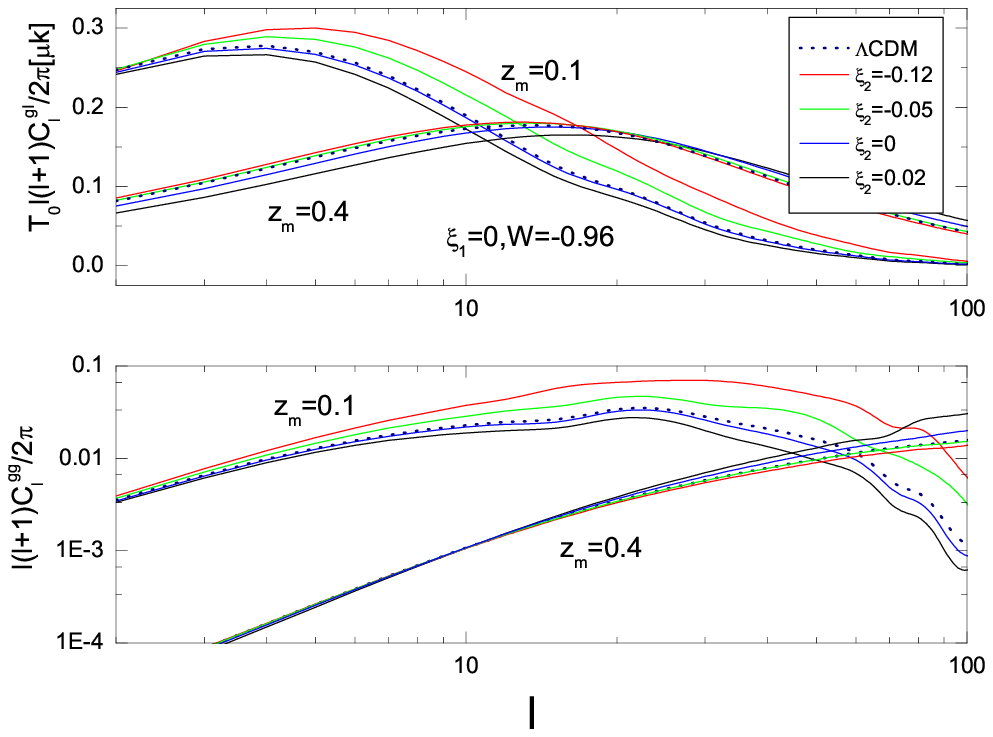}
\nonumber \\
    (c)&(d)\nonumber
  \end{tabular}
\end{center}
\caption{The small l CMB angular power spectra when the coupling is
proportional to the energy density of DE. The EoS of DE satisfies
$w>-1$. The up panel of (d) shows the cross-spectra and the lower
panel shows the galaxy power spectra.}\label{figone}
\end{figure}

When we choose the coupling between dark sectors proportional to the
energy density of DE, the perturbation was found stable with an EoS
$w>-1$ as well as $w<-1$\cite{22}. For chosen constant EoS
satisfying $w>-1$ and $C_e^2=1$, the result of large scale CMB
anisotropies for different couplings are shown in Fig~\ref{figone}a.
The lower set of dashed lines describes the late ISW effect, the
middle dotted lines show the combination of the SW effect together
with the early ISW effect and the top solid lines are for the total
large scale CMB angular power spectra. It is clear to see that such
kind of interaction between dark sectors leaves obvious signature in
the late ISW effect which influences the large scale total CMB
spectrum, but does not change much of the SW and early ISW effects.
Comparing to the non-coupling case ($\xi_2=0$), the positive
coupling with energy decay from DE to DM enhances the late ISW
effect, while the more negative coupling with more energy decay from
DM to DE suppresses more the late ISW contribution. This can be
understood from the evolution of the ISW source term, the change of
the gravitational potential displayed in Fig~\ref{figone}b. The
bigger value of the coupling results in the further change of the
gravitational potential. In Fig~\ref{figone}c we have shown the
influence of the sound speed. It can be seen that the bigger sound
speed of the DE makes the difference caused by different couplings
bigger in the large scale CMB power spectra.

\begin{figure}
\begin{center}
  \begin{tabular}{cc}
\includegraphics[width=3.2in,height=3.2in]{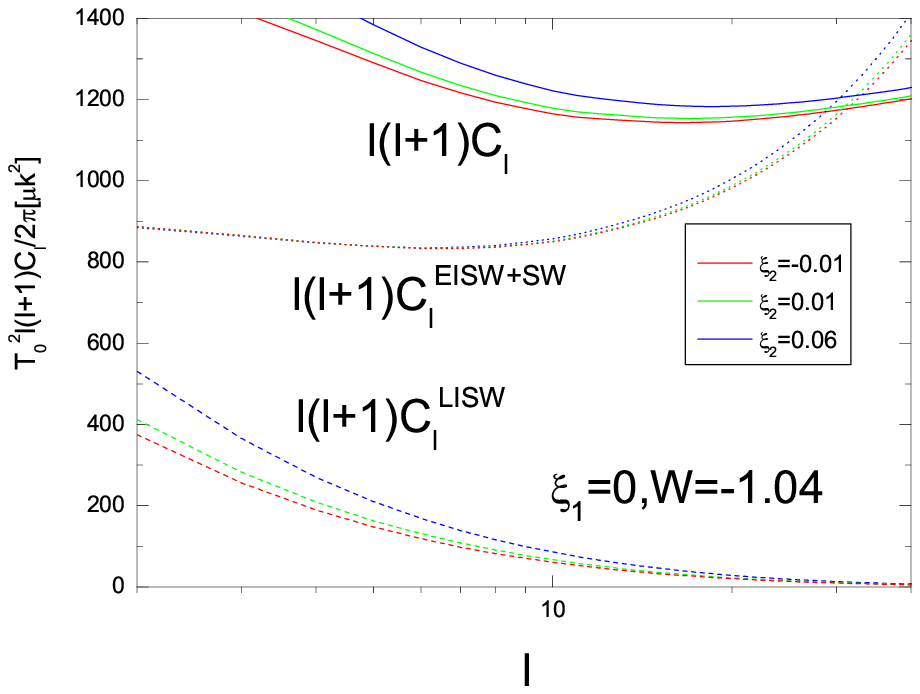}&
\includegraphics[width=3.2in,height=3.2in]{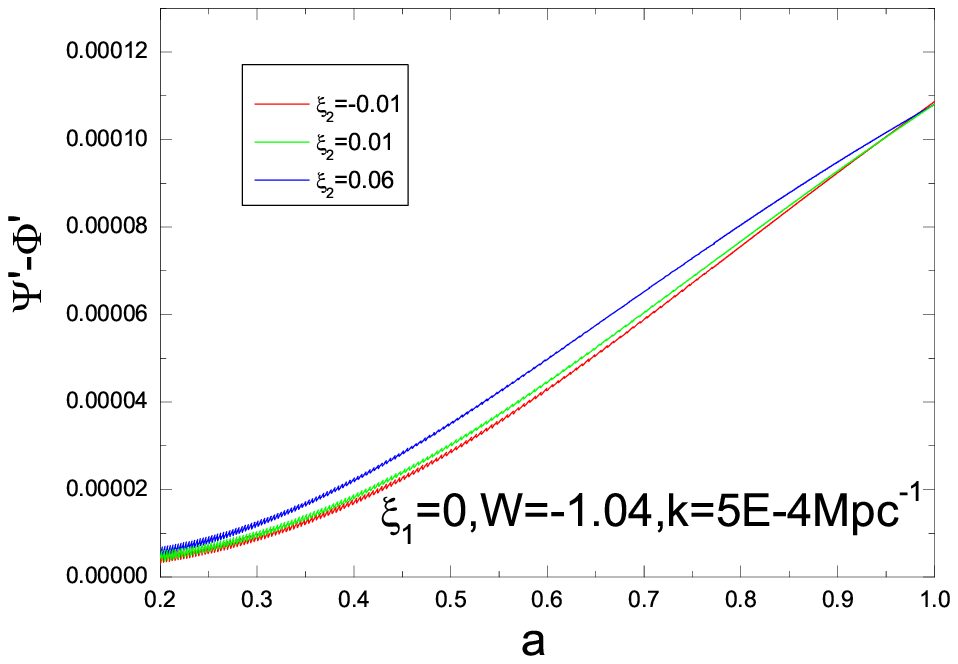}\nonumber \\
    (a)&(b)\nonumber \\
\includegraphics[width=3.2in,height=3.2in]{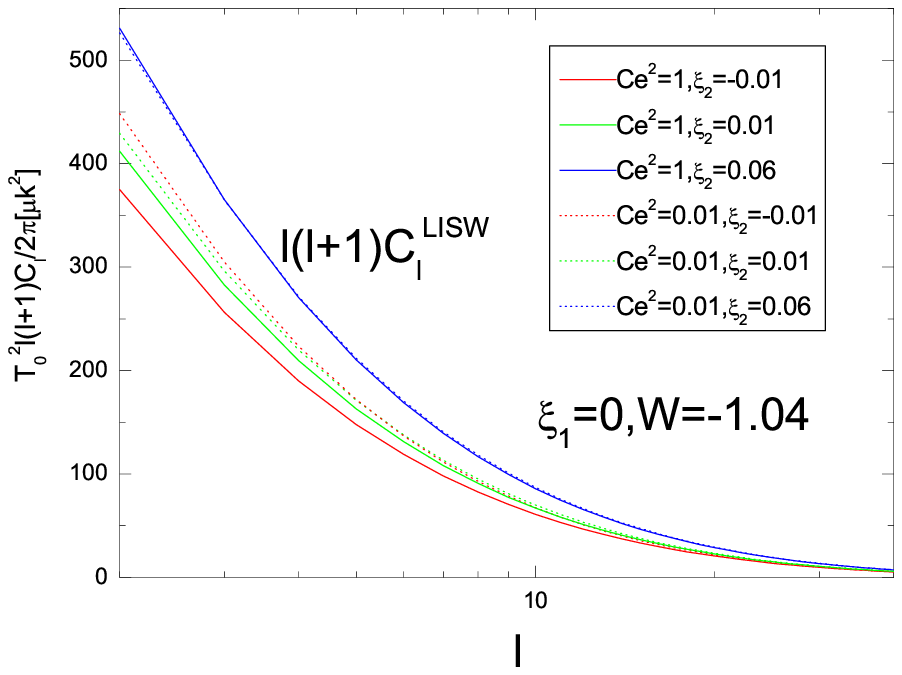}&\includegraphics[width=3.2in,height=3.2in]{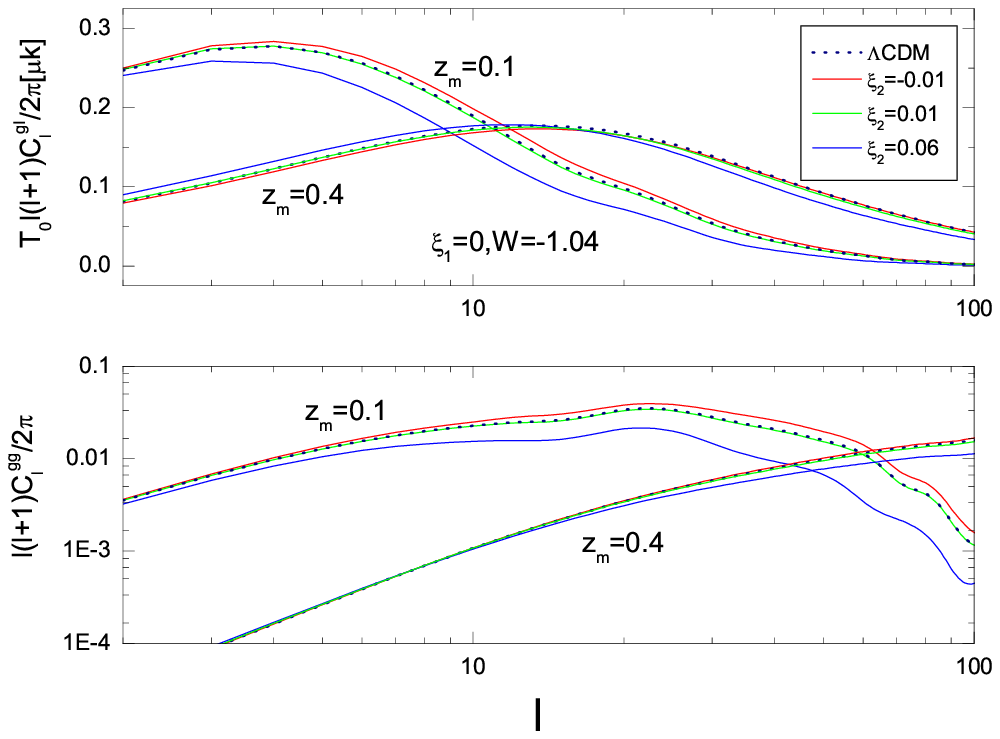}
\nonumber \\
    (c)&(d)\nonumber
  \end{tabular}
\end{center}
\caption{The small l CMB angular power spectra when the coupling is
proportional to the energy density of DE. The EoS of DE satisfies
$w<-1$.The up panel of (d) shows the cross-spectra and the lower
panel shows the galaxy power spectra.}\label{figtwo}
\end{figure}

\begin{figure}
\begin{center}
  \begin{tabular}{cc}
\includegraphics[width=3.2in,height=3.2in]{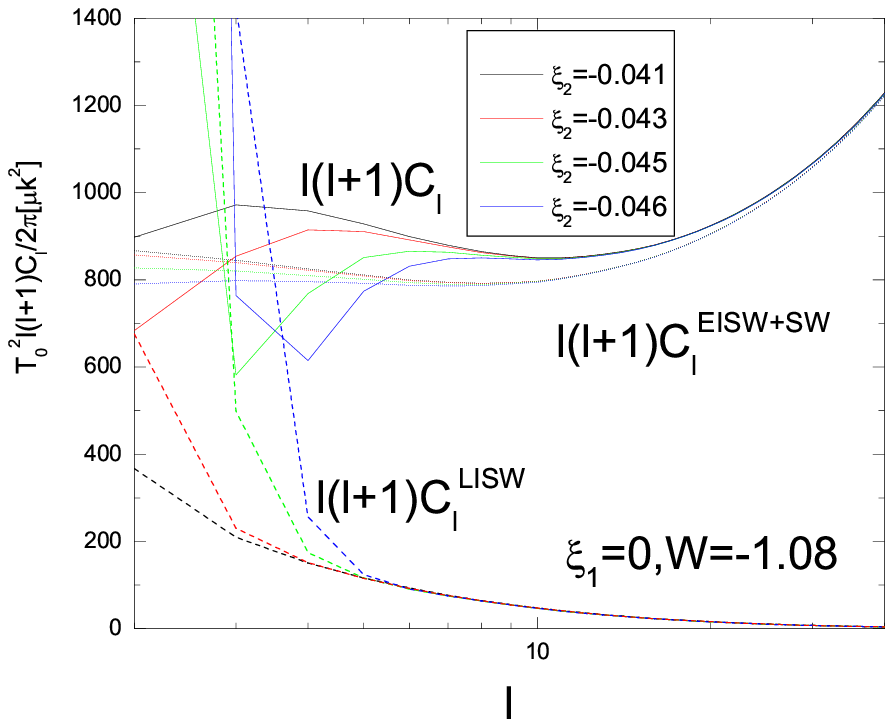}&
\includegraphics[width=3.2in,height=3.2in]{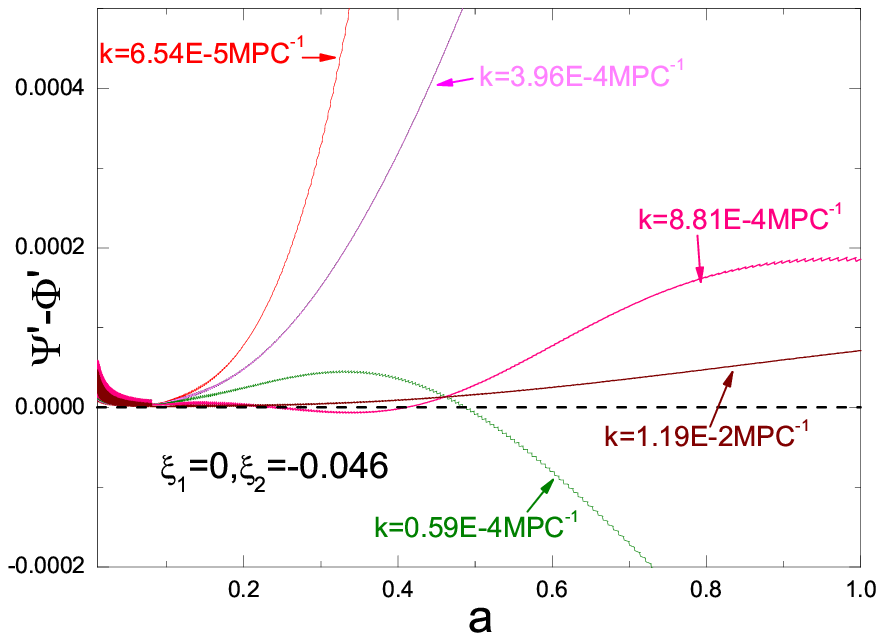}\nonumber \\
    (a)&(b)\nonumber \\
\includegraphics[width=3.2in,height=3.2in]{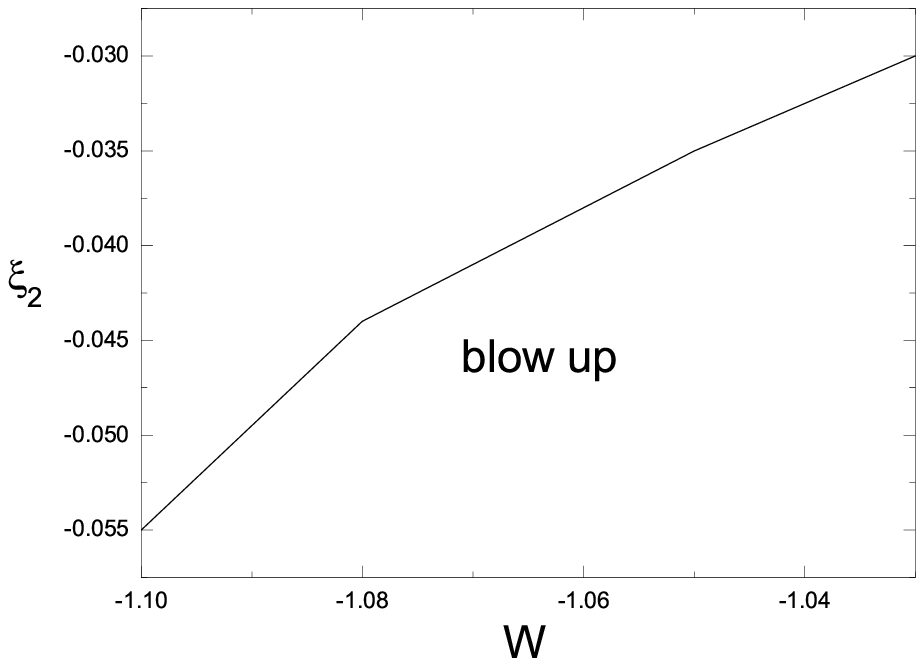}&\\
    (c)& \nonumber \\
  \end{tabular}
\end{center}
\caption{The irregular behaviors of the small l angular power
spectra when the couplings get too negative. Figure (c) indicates
the blow up with regard to $\xi_2$ and $w$.}\label{abnomal}
\end{figure}

Taking constant EoS satisfying $w<-1$ and $C_e^2=1$, the influence
of the interaction between dark sectors proportional to the energy
density of DE on the large scale CMB power spectra is shown in
Fig~\ref{figtwo}a. Again the coupling between dark sectors just
influences the late ISW effect and thus contributes to the large
scale CMB. Similar to that found in the $w>-1$ case, the late ISW
contribution becomes bigger as the coupling is more positive, which
is supported by the correlated potential behavior in
Fig~\ref{figtwo}b.  The difference in the large scale CMB power
spectra induced by this kind of interaction reduces again for
smaller sound speed of DE, see Fig~\ref{figtwo}c. In the case
$w<-1$, we observed some irregular phenomenons when the coupling is
more negative. As shown in Fig~\ref{abnomal}a, there appears a blow
up behavior in the late ISW contribution for very small $l$ and so
does the very large scale total power spectra. For more negative
coupling, this blow up appears because when the spatial wavenumber
$k$ becomes small enough there is a sharp increase in the potential
as shown in Fig~\ref{abnomal}b. The relation between the negative
coupling leading to the blow up with EoS of DE is shown in
Fig~\ref{abnomal}c. Since observations suggest the deficit of large
scale CMB power, thus we can rule out the too negative coupling in
this interaction form.


\begin{figure}
\begin{center}
  \begin{tabular}{cc}
\includegraphics[width=3.2in,height=3.2in]{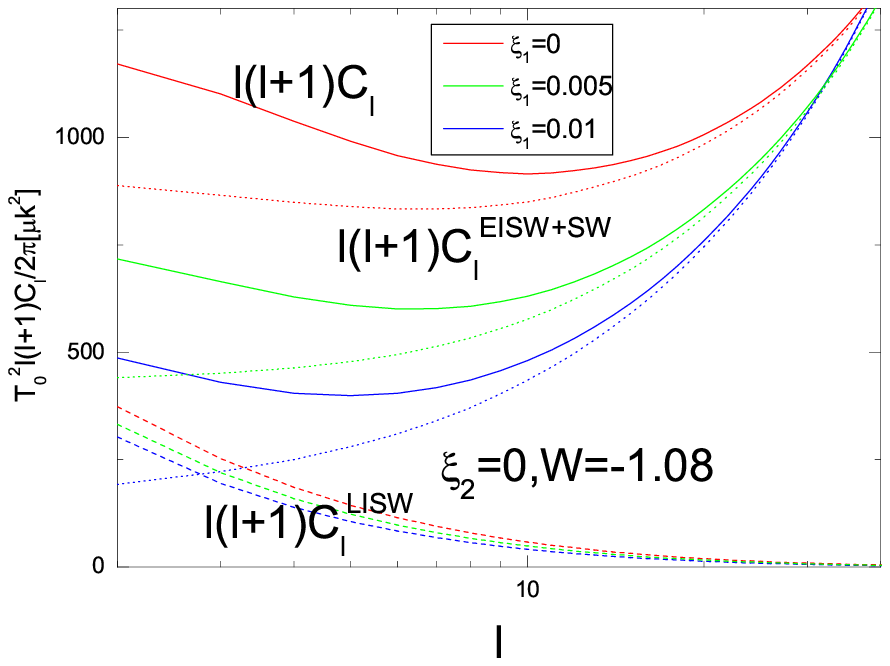}&
\includegraphics[width=3.2in,height=3.2in]{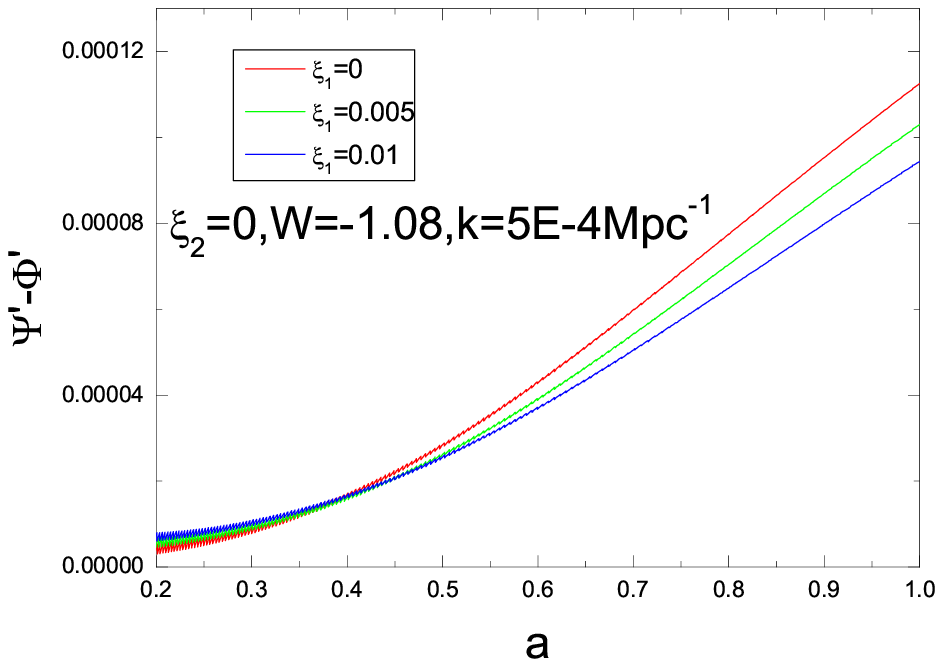}\nonumber \\
    (a)&(b)\nonumber \\
\includegraphics[width=3.2in,height=3.2in]{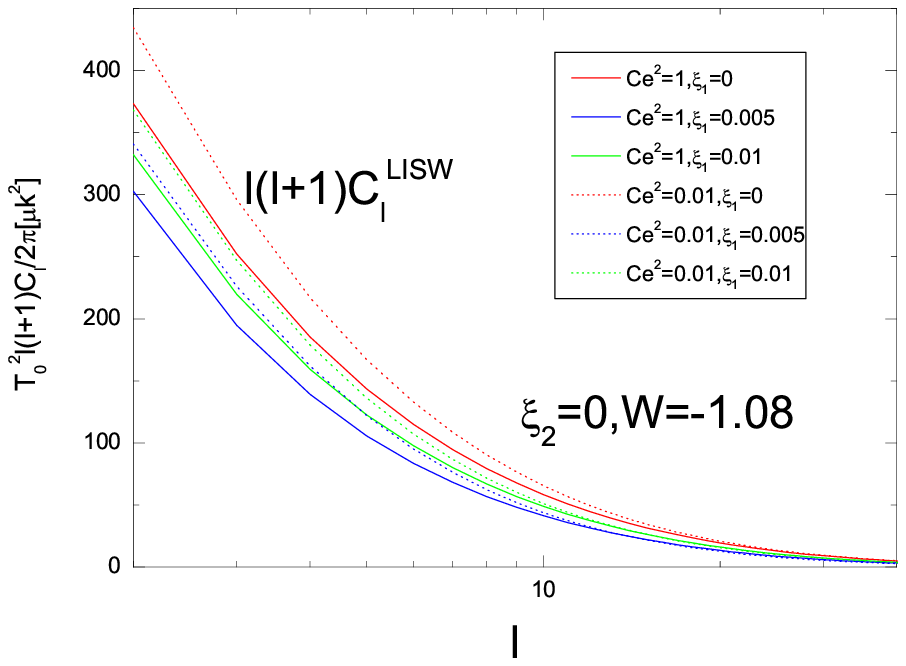}&\includegraphics[width=3.2in,height=3.2in]{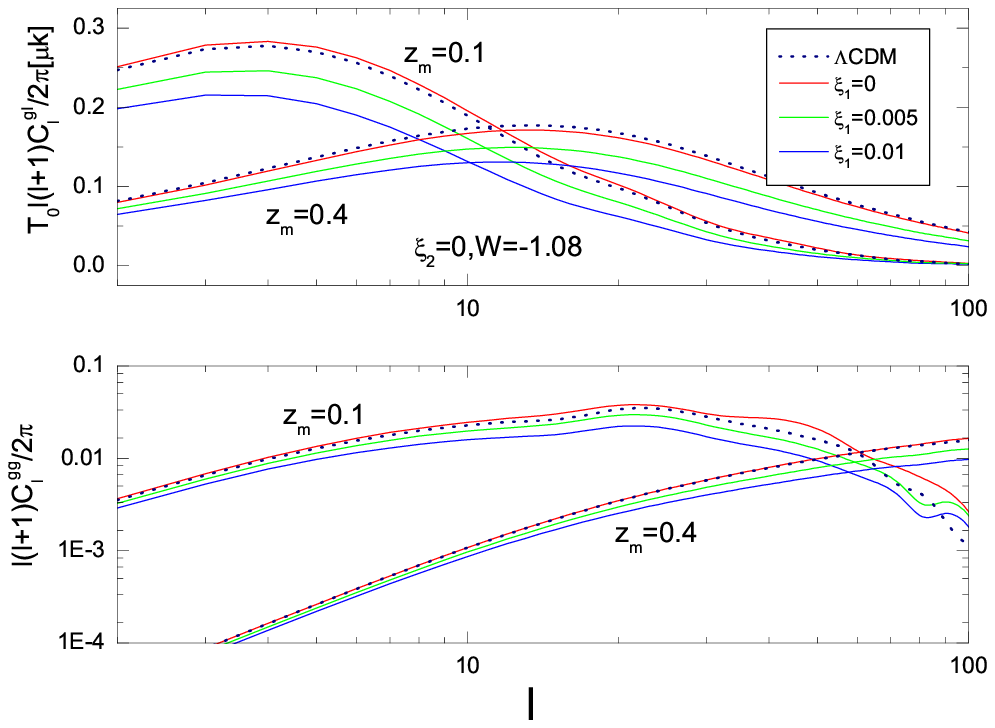}
\nonumber \\
    (c)&(d)\nonumber
  \end{tabular}
\end{center}
\caption{The small l CMB angular power spectra when the coupling is
proportional to the energy density of DM. The DE EoS satisfies
$w<-1$. The up panel of (d) shows the cross-spectra and the lower
panel shows the galaxy power spectra.}\label{figthree}
\end{figure}

Now we turn to discuss the coupling between dark sectors
proportional to the energy density of DM, the perturbation was found
stable with an EoS $w<-1$ only\cite{22}. Choosing constant EoS with
$w<-1$ and sound speed $C_e^2=1$, the influence of the interaction
between DE and DM is displayed in Fig~\ref{figthree}a. We see that
for this kind of phenomenological coupling, the SW effect together
with the early ISW effect get more modification than the late ISW
effect due to the coupling. Contrary to the coupling proportional to
the energy density of DE case, we found that for this kind of
interaction the more positive coupling results in the more
suppression in the CMB anisotropies. This is reasonable if we look
at its correlated potential behavior, see Fig~\ref{figthree}b. In
Fig~\ref{figthree}c we have shown that the smaller sound speed of DE
will enhance the difference in the very large scale power in the
temperature map, which is also different from the result we have
seen above for the coupling proportional to the DE energy density.

\begin{figure}
\begin{center}
  \begin{tabular}{cc}
\includegraphics[width=3.2in,height=3.2in]{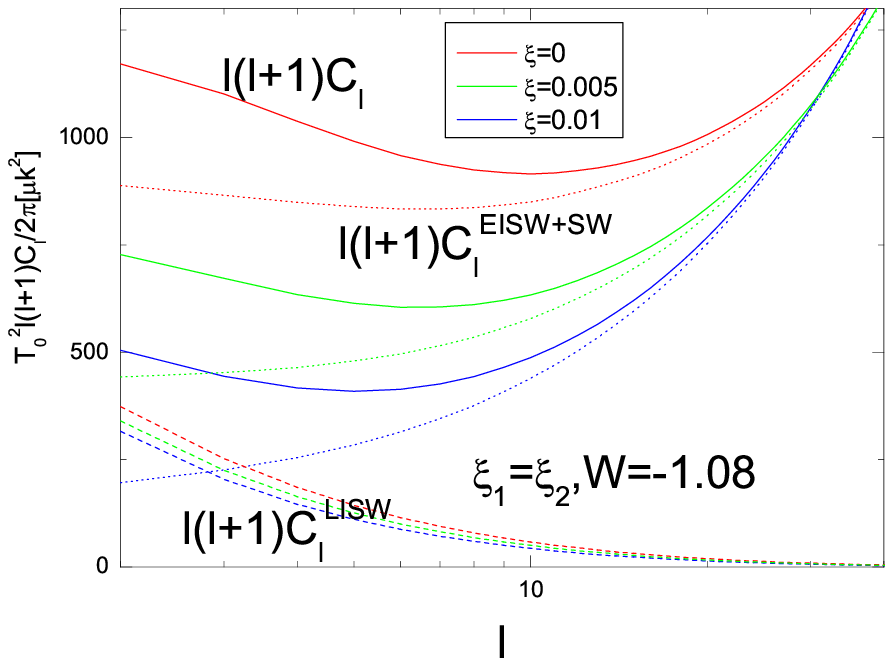}&
\includegraphics[width=3.2in,height=3.2in]{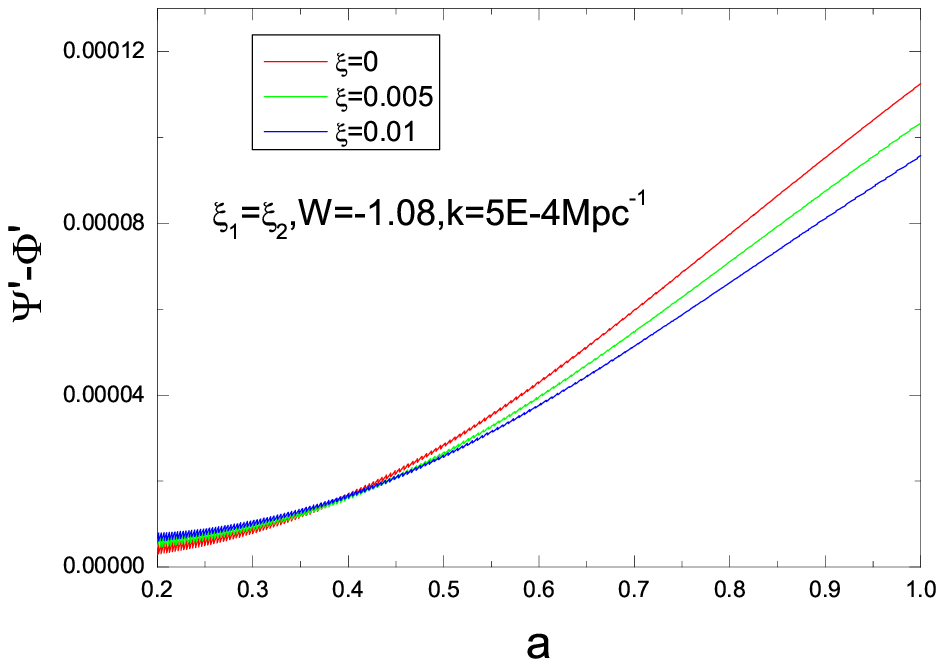}\nonumber \\
    (a)&(b)\nonumber \\
\includegraphics[width=3.2in,height=3.2in]{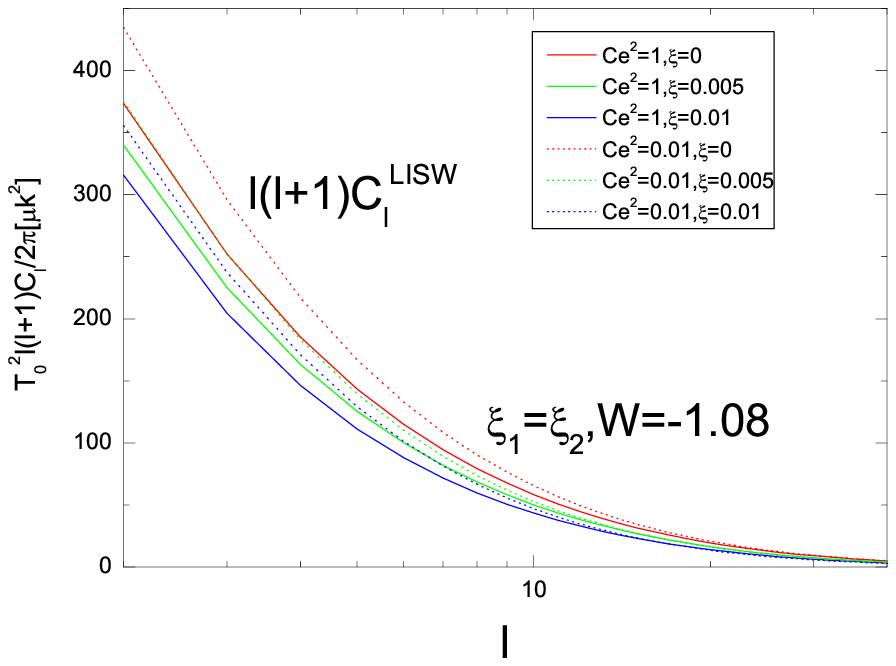}&\includegraphics[width=3.2in,height=3.2in]{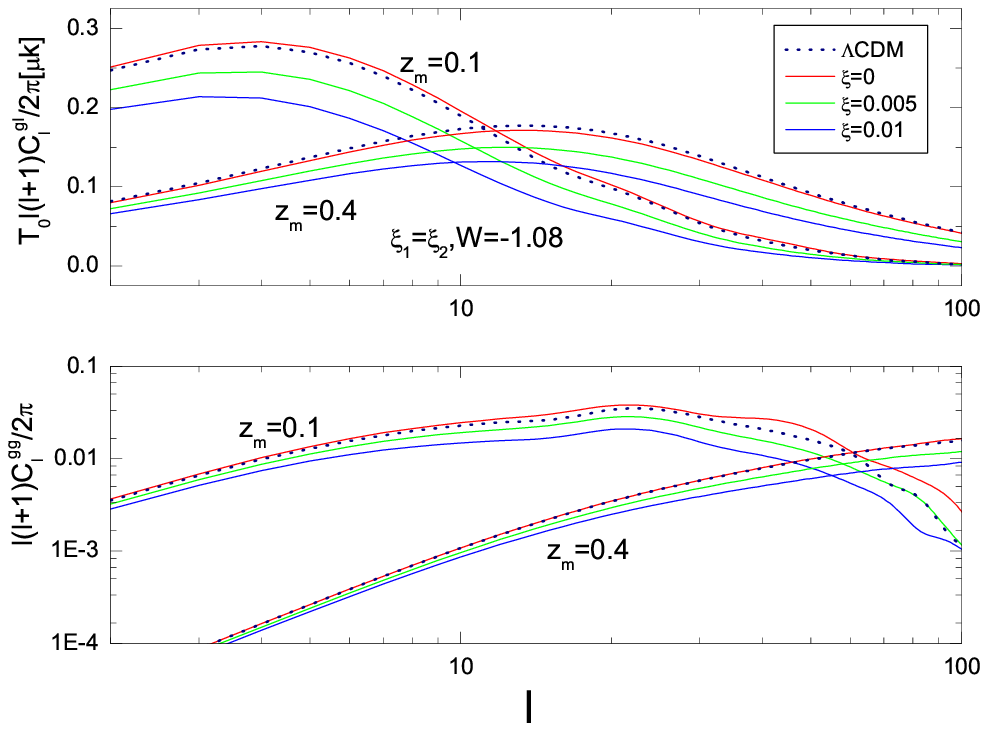}
\nonumber \\
    (c)&(d)\nonumber
  \end{tabular}
\end{center}
\caption{The small l CMB angular power spectra when the coupling is
proportional to the energy density of total dark sectors and the EoS
$w<-1$.The up panel of (d) shows the cross-spectra and the lower
panel shows the galaxy power spectra.}\label{total}
\end{figure}

For the interaction between DE and DM  proportional to the energy
density of total dark sectors, we list our results in
Fig~\ref{total}, which is very similar to that we observed when the
interaction proportional to the energy density of DM.

With the formalism of the perturbation theory when DE and DM are in
interaction, we have analyzed the signature of the coupling between
dark sectors from the large scale CMB anisotropies. The large scale
power in the temperature map looks very different when interactions
are included and this provides the possibilities to use the large
scale CMB information to constrain the coupling between dark
sectors.

By far we only discuss the influence of the ISW effect on the CMB auto power
spectrum. Since the late ISW effect arises from the time varying gravitational
potential, which correlates with the large scale structure (LSS) of the
universe, the cross correlating CMB with tracers of
LSS provides another way of measuring the ISW effect
\citep{ISWcross}.  Progresses in CMB and LSS surveys have enabled detections
of the ISW-LSS cross correlation at $\sim 3\sigma$ level
\citep{ISWmeasurements}. The S/N can be further improved by a factor of a few
for future all sky LSS surveys.  This cross correlation technique has a
number of advantages. First, the primary CMB does not correlate with the LSS
and thus does not bias the ISW measurement. Second, the cross
correlation signal is  $\propto \Delta T_{\rm ISW}$, which tells whether the
potential decays or grows. The later case can happen in our
interacting dark matter-dark energy models (e.g, panel b, Fig. \ref{abnomal})
and
in modified gravity models. As a comparison, the CMB auto power spectrum is
only sensitive to $\Delta T^2_{\rm ISW}$ and thus loses this
capability. Third, through the redshift distribution of the LSS, we are able
to recover the redshift evolution of the gravitational potential and thus
infer more details on the dark matter-dark energy interaction. Thus the
ISW-LSS cross correlation is potentially powerful for probing the dark
matter-dark energy interaction. However, due to the low
signal-to-noise ratio of the current ISW-LSS measurements and the complexities
in  the theoretical interpretation (e.g. the galaxy bias), we will not confront
our model predictions against the existing ISW-LSS cross correlation
measurement in this paper. Instead, we will just calculate the expected cross
correlation signal between the ISW effect and galaxies, for some representive
cases.

 The auto- and cross-correlation power spectra are given by
\begin{eqnarray}
C_l^{gg}  =  4\pi \int \frac{dk}{k}
\mathcal{P}_{\chi}(k)I_l^{g*}(k)I_l^g(k)\\
C_l^{gI}  =  4\pi \int \frac{dk}{k}
\mathcal{P}_{\chi}(k)I_l^{g*}(k)\Delta_l^{ISW}(k),
\end{eqnarray}
where the integrand for galaxy densities $I_l^{g}(k)$ reads,
\begin{equation}
I_l^{g}(k)= \int dz b_g(z)\Pi(z)(D_{gc}+D_{gb})j_l[k\chi(z)].
\end{equation}
Here $b_g(z)$ is the galaxy bias, $\Pi(z)$is the redshift
distribution and $\chi(z)$ is the conformal distance, or
equivalently the look-back time from the observer at redshift $z=0$,
\begin{equation}
\chi(z) = \int_0^z \frac{dz'}{H(z)} = \int_{\tau_i(z)}^{\tau_0}d\tau
= \tau_0 - \tau_i(z).
\end{equation}
We assume $b(z)\sim1$ for simplicity and adopt the redshift
distribution of the form\cite{select},
\begin{equation}
\Pi(z) = \frac{3}{2} \frac{z^2}{z_0^3}{\rm exp}\left[
-(\frac{z}{z_0})^{3/2}\right]
\end{equation}
normalized to unity and peaking near the median redshift
$z_m=1.4z_0$. For illustrative purpose,we choose $z_m=0.1$ and
$z_m=0.4$ throughout our analysis. The first choice resembles a
shallow survey like 2MASS and the second one resembles a survey
similar to SDSS photo-z galaxy samples.

 When the coupling is proportional to the energy density of DE
and the EoS is larger than $-1$($w>-1$), the cross power spectra and
auto correlation power spectra of galaxies are shown in
Fig~\ref{figone}d. For lower redshift galaxies survey $z_m=0.1$,
comparing with the LCDM model we see that the couplings
significantly change both the cross spectra and the auto spectra.
The negative couplings enhance the power of the correlation while
the positive couplings hinder such correlation. When the EoS is
smaller than $-1$, we find very similar results as shown in
Fig~\ref{figtwo}d. However, for deeper redshift galaxies survey $z_m=0.4$,
the couplings do not imprint significantly on the cross power
spectra and auto correlation power spectra of galaxies as compared
with $z_m=0.1$.

When the coupling is proportional to the energy density of DM or
total dark sectors, we find from
Fig~\ref{figthree}d,Fig~\ref{total}d that the cross power spectra
are more sensitive to the couplings at lower $l$ part than that of
higher $l$ part when the median redshift around $0.1$. This feature
is different from that shown in Fig~\ref{figone}d and
Fig~\ref{figtwo}d when the interaction is proportional to the energy
density of DE, where it was found that at small $l$ when ISW effect
amplified, the ISW-LSS cross-correlation is not so much different
due to the coupling. While in the other way, for higher redshift survey $z_m=0.4$, we find less such effect on the cross and auto spectra.

The qualitative behaviors presented here show that the 2MASS survey
has more possibility in discriminating the interaction between dark
sectors than that of SDSS survey. It is expected in the future that
galaxy surveys with photo-z measurements or even spec-z
measurements, along with better CMB measurements, could provide
better ISW-LSS cross-correlation measurements at each redshift bin.
However, since the latest result on the measurement of this effect
is not better than $\sim 3\sigma$ \cite{Giannantonio} in accuracy,
we will not incorporate the relevant data set in our global fitting
in this work.

In the next section we are going to report the fitting result by
comparing with observations of CMB.

\section{constraints using observational data and the coincidence problem}

In order to further see the signature of the interaction between DE
and DM, in this section we will compare our model with observational
data by using joint likelihood analysis. We take the parameter space
as
$$P=(h,\omega_b,\omega_{cdm},\tau,\ln[10^{10}A_s],n_s,\xi_1,\xi_2,w)$$
where $h$ is the hubble constant,
 $\omega_b=\Omega_bh^2, \omega_{cdm}=\Omega_{cdm}h^2$, $A_s$ is the
amplitude of the primordial curvature perturbation, $n_s$ is the
scalar spectral index, $\xi_1$ and $\xi_2$ are coupling constants
proportional to the energy density of DM and DE respectively, $w$ is
the EoS of DE. We choose the flat universe with $\Omega_k=0$ and our
work is based on CMBEASY code\cite{easy}.

In the global fitting, we have used CMB data coming from WMAP5
temperature and polarization power spectra. We used Gibbs sampling
routine provided by WMAP team for the likelihood calculation. In the
small scale CMB measurements, we included BOOMERanG\cite{BOOMERanG},
CBI \cite{CBI}, VSA\cite{VSA} and ACBAR\cite{ACBAR} data. In order
to get better constraint on the background evolution, we have added
SNIa\cite{SNeIa} data and marginalized over the nuisance parameters.
We also incorporated the data from large scale luminous red
galaxies(LRGs) survey, we used SDSS\cite{SDSS} data as powerful
constraint on real-space power spectrum $P(k)$ at redshift
$z\sim0.1$.

Since the terms incorporating $Q_d^0$ on the left hand sides of
eq.(\ref{density},\ref{velocity}) encounter the irreducible
singularity when $w$ crossing $-1$, we therefore set prior on EoS
either $w>-1$ or $w<-1$. We take the effective sound speed $C_e^2=1$
in our analysis.

The global fitting results for different forms of interaction
between dark sectors are shown below:

\subsection{The interaction proportional to DE energy density}
For the DE EoS $w>-1$, we have the result
\begin{center}
\begin{tabular}{|c|c|c|c|c|c|c|c|c|}
\hline
   $h$ & $\Omega_bh^2$ & $\Omega_{cdm}h^2$ & $\tau$ & $ \ln[10^{10}A_s] $ & $n_s$ & $\xi_1=0,\xi_2\neq 0$ & $1+w>0$  \\
\hline
  $0.687_{ -0.013}^{+0.013}$ & $0.0225_{-0.0005}^{+0.0006}$ & $0.107_{-0.009}^{+0.007}$ & $0.091_{-0.016}^{+0.017}$ & $3.109_{-0.031}^{+0.033}$ & $0.963_{-0.013}^{+0.013}$ & $-0.028_{-0.032}^{+0.023}$ & $<0.052$   \\
\hline
\end{tabular}
\end{center}

For the phantom DE EoS $w<-1$,we constrain $\xi_2>0.030$ to avoid
the blow-up and we have
\begin{center}
\begin{tabular}{|c|c|c|c|c|c|c|c|c}
\hline
   $h$ & $\Omega_bh^2$ & $\Omega_{cdm}h^2$ & $\tau$ & $ \ln[10^{10}A_s] $ & $n_s$ & $\xi_1=0,\xi_2\neq 0$ & $1+w<0$\\
\hline
   $0.700_{ -0.011}^{+0.012}$ & $0.0223_{-0.0005}^{+0.0006}$ & $0.119_{-0.006}^{+0.009}$ & $0.086_{-0.016}^{+0.019}$ & $3.105_{-0.032}^{+0.033}$ & $0.956_{-0.014}^{+0.013}$ & $-0.010_{-0.020}^{+0.025}$ & $-0.051_{-0.043}^{+0.051}$ \\
\hline
\end{tabular}
\end{center}

\begin{figure}
\includegraphics[width=2.8in,height=2.8in]{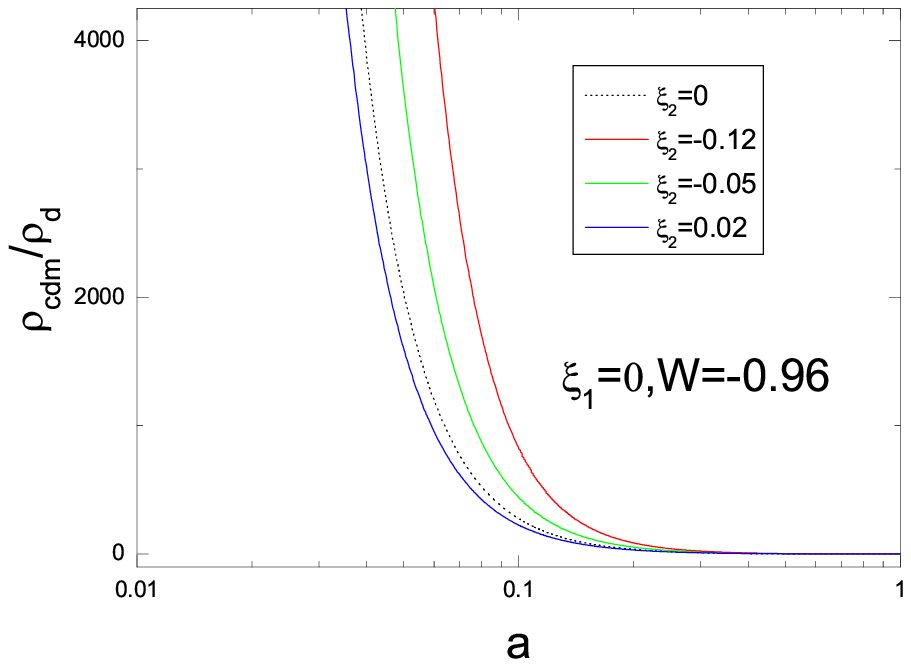}
\includegraphics[width=2.8in,height=2.8in]{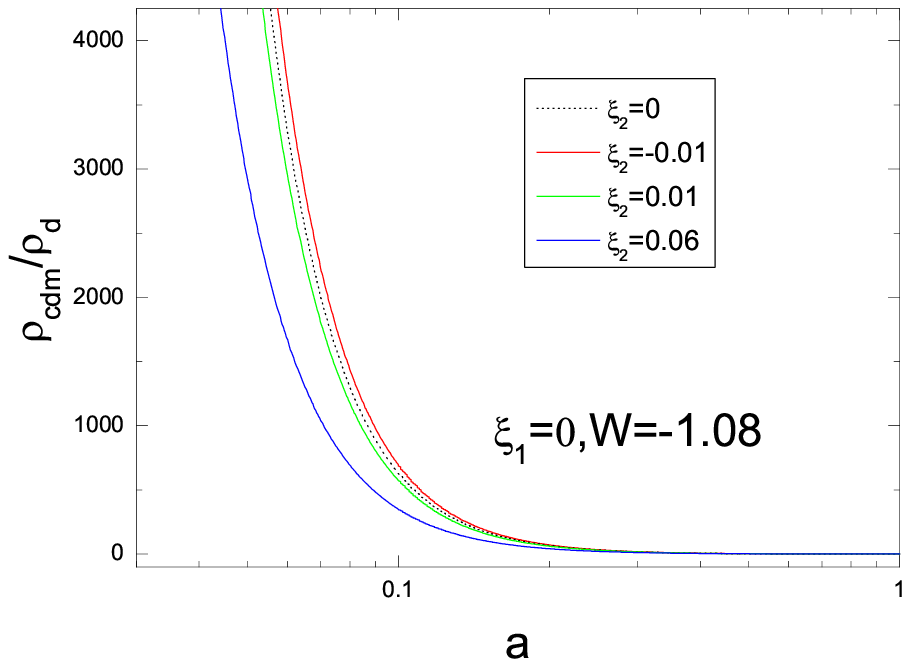}
\caption{The behaviors of the ratio $r=\rho_{c}/\rho_{d}$ when the
coupling is proportional to the energy density of DE. The left one
for $w>-1$ and the right one for $w<-1$. }\label{five}
\end{figure}

In both cases we observed from the global fitting that the coupling
between DE and DM in the $1\sigma$ range can either be positive or
negative. Using the best-fit results, we have studied the
coincidence problem. We paid attention to the ratio of energy
densities between DE and DM, $r = \rho_{c}/\rho_{d}$, and its
evolution. No matter for $w>-1$ or $w<-1$, we observed that a slower
change of $r$ for positive coupling as compared to the
noninteracting case. This means that the period when energy
densities of DE and DM are comparable is longer when there is a
positive coupling between DE and DM, see Fig~\ref{five} for an
example.

\subsection{The interaction proportional to DM energy density}
Now we report the fitting result for the interaction between dark
sectors proportional to the energy density of DM.
\begin{center}
\begin{tabular}{|c|c|c|c|c|c|c|c|c}
\hline
   $h$ & $\Omega_bh^2$ & $\Omega_{cdm}h^2$ & $\tau$ & $ \ln[10^{10}A_s] $ & $n_s$ & $\xi_1>0,\xi_2=0$ & $1+w<0$\\
\hline
   $0.690_{ -0.015}^{+0.015}$ & $0.0224_{-0.0005}^{+0.0005}$ & $0.121_{-0.003}^{+0.003}$ & $0.094_{-0.017}^{+0.016}$ & $3.115_{-0.033}^{+0.034}$ & $0.953_{-0.013}^{+0.013}$ & $0.0007_{-0.0006}^{+0.0006}$ & $-0.072_{-0.053}^{+0.072}$ \\
\hline
\end{tabular}
\end{center}

\begin{figure}
\begin{center}
  \begin{tabular}{cc}
\includegraphics[width=2in,height=2in]{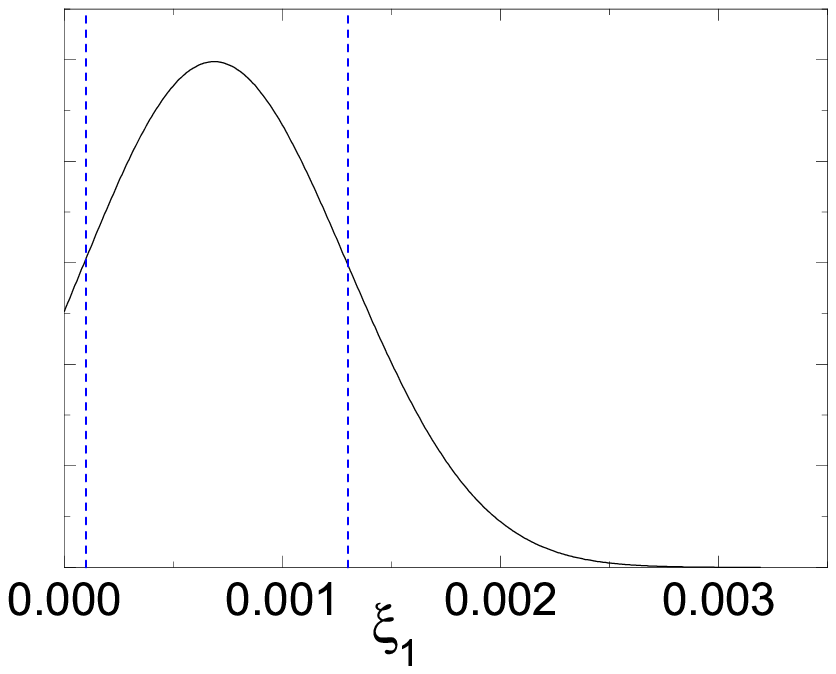}&
\includegraphics[width=2in,height=2in]{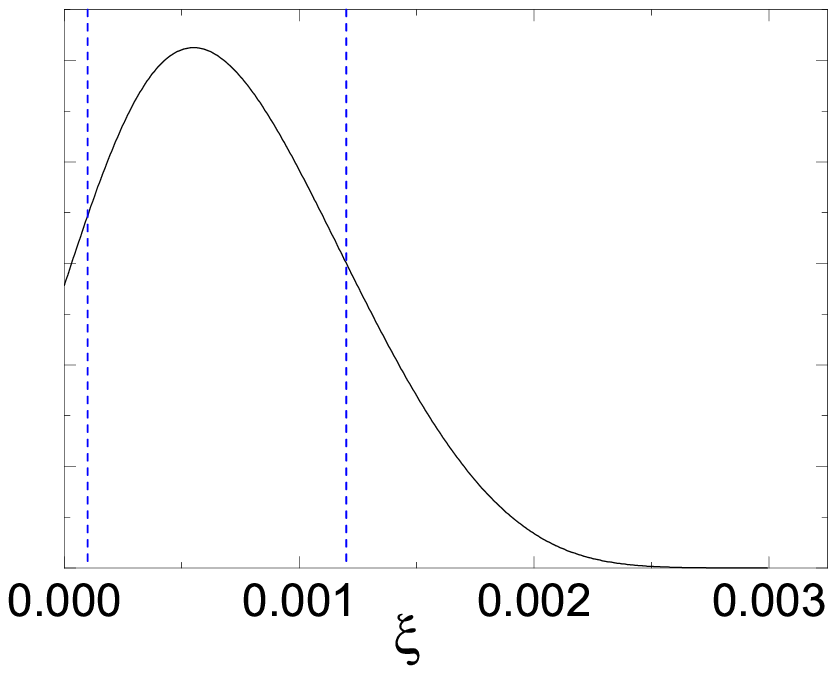}\nonumber \\
    (a)&(b)\nonumber \\
  \end{tabular}
\end{center}
\caption{The likelihood for the coupling constant when the
interaction is proportionl to the energy density of DM (left) and
the total dark sectors (right).  }\label{DM}
\end{figure}

\begin{figure}
\includegraphics[width=2.8in,height=2.8in]{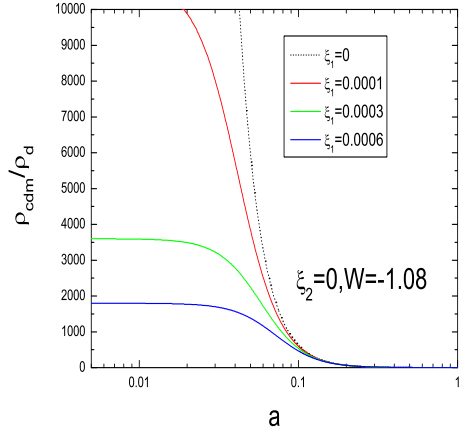}
\caption{The behaviors of the ratio $r=\rho_{cdm}/\rho_{d}$ when the
coupling is proportional to the energy density of DM. }\label{seven}
\end{figure}

It is interesting to see that for this kind of interaction, the
global fitting told us that in $1\sigma$ range the coupling between
DE and DM is always positive, see Fig~\ref{DM}a. This is
encouraging, since for positive coupling with the energy decay from
DE to DM we can always have the slower change of $r$ as compared to
the noninteracting case as shown in Fig~\ref{seven}. Thus with this
kind of interaction the coincidence problem can be overcome. Another
interesting point is that employing this kind of interaction form we
can constrain the coupling to a very precise value compared with the
interaction proportional to the energy density of DE. This is
because this kind of interaction shows up not only in the late ISW
effect, but also the SW and early ISW effects as shown in
Fig~\ref{figthree}. As displayed in Fig~\ref{seven}, DE and DM can
trace each other from very early period. It can leave imprints in at
smaller scales (bigger $l$ multipoles) where there are more
independnet modes to improve observation accuracy. This can help to
reduce the uncertainty in determining the coupling constant for this
kind of interaction.

\subsection{The interaction proportional to energy density of total dark sectors}
For the interaction proportional to the energy density of total dark
sectors, we have obtained the similar result as in the above
subsection. The results are shown below:
\begin{center}
\begin{tabular}{|c|c|c|c|c|c|c|c|c}
\hline
   $h$ & $\Omega_bh^2$ & $\Omega_{cdm}h^2$ & $\tau$ & $ \ln[10^{10}A_s] $ & $n_s$ & $\xi=\xi_1=\xi_2>0$ & $1+w<0$\\
\hline
   $0.690_{ -0.014}^{+0.014}$ & $0.0224_{-0.0006}^{+0.0006}$ & $0.121_{-0.003}^{+0.004}$ & $0.093_{-0.017}^{+0.018}$ & $3.114_{-0.033}^{+0.036}$ & $0.955_{-0.014}^{+0.014}$ & $0.0006_{-0.0005}^{+0.0006}$ & $-0.065_{-0.054}^{+0.065}$ \\
\hline
\end{tabular}
\end{center}

\begin{figure}
\includegraphics[width=2.8in,height=2.8in]{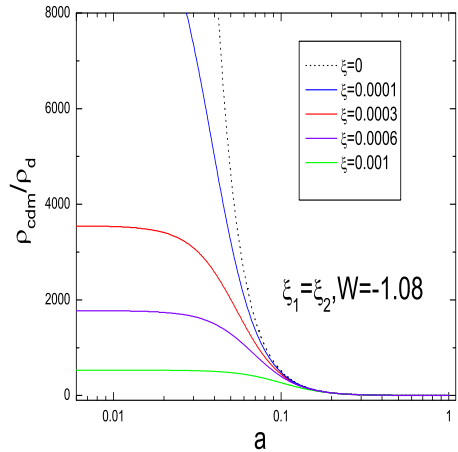}
\caption{The behaviors of the ratio $r=\rho_{cdm}/\rho_{d}$ when the
coupling is proportional to the energy density of total dark
sectors.}\label{eight}
\end{figure}

The likelihood of the coupling constant is shown in Fig~\ref{DM}b.
The evolution of the ratio of energy densities between DE and DM for
this kind of interaction is shown in Fig~\ref{eight}.

\section{conclusions and discussions}

We have reviewed the formalism of the perturbation theory when there
is interaction between DE and DM. Based upon this formalism we have
studied the signature of the interaction between dark sectors in the
CMB large scale temperature fluctuations. We found that in addition
to disclose the DE EoS, sound speed, the late ISW effect is a
promising tool to measure the coupling between dark sectors. When
the interaction between DE and DM takes the form proportional to the
energy density of DM and the total dark sectors, because in these
cases DE and DM started to chase each other since early time, the
interaction not only presents in the late ISW source term but also
leaves imprint in the SW and early ISW effects. These properties
provide a possible way to examine the interaction between DE and DM
even from  smaller scale in  CMB observations.

 We have performed the global fitting by using the CMB power spectrum
data including WMAP5 data and balloon observational data together
with SNIa and SDSS data to constrain the interaction between DE and
DM. When the interaction between DE and DM takes the form
proportional to the energy density of DM and the total dark sectors,
since it leaves more information in the CMB power spectra, not only
just in the very large scale, the coupling can be constrained in a
very precise range. In $1\sigma$ the coupling is positive indicating
that there is energy transfer from DE to DM. This kind of energy
transfer can help to alleviate the coincidence problem compared to
the noninteracting case.

It is of great interest to extend our study to a field theory
description of the interaction between DE and DM and examine its
signature in the large scale CMB power spectra. A possible field
theory model was proposed in \cite{field} and further investigation
in this direction is called for.

\acknowledgements{ This work has been supported partially by NNSF of
China, Shanghai Science and Technology Commission and Shanghai
Education Commission. We would like to acknowledge helpful
discussions with X. L. Chen, Y. G. Gong, F. Q. Wu and the
hospitality of KITPC where main part of the work was carried out.}


\begin{thebibliography}{99}
%%%%%%%%%%%%%%%%%%%%%%%%%%%%%%%%%%%%%%%%%%%%%%%%%%%%%%%%%%%%%%%%%%%%%%%%%

\bibitem{1} S. J. Perlmutter et al., Nature 391, 51 (1998); A. G. Riess et al., Astron. J. 116, 1009 (1998); S. J. Perlmutter et al.,
Astroph. J. 517, 565 (1999); J. L. Tonry et al., Astroph. J. 594, 1,
(2003); A. G. Riess et al., Astroph. J. 607, 665 (2005); P. Astier
et al., Astron. Astroph. 447, 31 (2005); A G. Riess et al., Astroph.
J. 659, 98 (2007).
\bibitem{2} M. Kowalski et al., 2008, arXiv:0804.4142.
\bibitem{3} M. Tegmark et al., Phys. Rev. D 74, 123507 (2006).
\bibitem{10} L. Amendola, Phys. Rev. D 62, 043511 (2000); L. Amendola and C. Quercellini, Phys. Rev. D 68, 023514 (2003); L.
Amendola, S. Tsujikawa and M. Sami, Phys. Lett. B 632, 155 (2006).
\bibitem{11} D. Pavon, W. Zimdahl, Phys. Lett. B 628, 206 (2005), S.
Campo, R. Herrera, D. Pavon, Phys. Rev.D 78, 021302(R) (2008).
\bibitem{12} G. Olivares, F. Atrio-Barandela and D. Pavon, Phys. Rev. D 74, 043521
(2006).
\bibitem{13} C. G. Boehmer, G. Caldera-Cabral, R. Lazkoz, R.
Maartens, Phys. Rev. D 78, 023505 (2008).
\bibitem{14} S. B. Chen, B. Wang,
J. L. Jing, Phys.Rev.D 78, 123503 (2008).
\bibitem{15} B. Wang, J. Zang,
C.-Y. Lin, E. Abdalla and S. Micheletti, Nucl. Phys. B 778, 69
(2007).
\bibitem{16} W. Zimdahl, Int. J. Mod. Phys. D 14, 2319 (2005).
\bibitem{17} D. Pavon, B. Wang, Gen. Relav. Grav. (in press), arXiv:0712.0565.
\bibitem{18} B. Wang, C.-Y. Lin, D. Pavon, E. Abdalla, Phys. Lett. B 662, 1,
(2008).
\bibitem{19} O. Bertolami, F. Gil Pedro and M. Le Delliou, Phys.
Lett. B 654, 165 (2007). O. Bertolami, F. Gil Pedro and M. Le
Delliou, arXiv:0705.3118v1.
\bibitem{20} Z. K. Guo, N. Ohta and S.
Tsujikawa, Phys. Rev. D 76, 023508 (2007).
\bibitem{21} J. Valiviita, E.
Majerotto, R. Maartens, JCAP 07, 020 (2008), ArXiv:0804.0232.
\bibitem{22} J.
H. He, B. Wang, E. Abdalla, Phys. Lett. B 671, 139 (2009),
ArXiv:0807.3471
\bibitem{23} W. Zimdahl, D. Pavon, L.P. Chimento, Phys.
Lett. B 521, 133 (2001); L.P. Chimento, A.S. Jakubi, D. Pavon, W.
Zimdahl, Phys. Rev. D 67, 083513 (2003).
\bibitem{24} E. Abdalla, L.Raul W.
Abramo, L. Sodre Jr., B. Wang, Phys. Lett. B673, 107 (2009)
ArXiv:astro-ph/0710.1198.
\bibitem{25} J.
H. He, B. Wang, JCAP 06, 010 (2008), ArXiv:0801.4233.
\bibitem{26} C. Feng,
B. Wang, E. Abdalla, R. K. Su, Phys. Lett. B 665, 111 (2008),
arXiv:0804.0110.
\bibitem{27} B. Jackson, A. Taylor, A. Berera,
arXiv:0901.3272.
\bibitem{28} A. Refrefier, et al, ArXiv: 0802.2522.
\bibitem{29} S.
Das, P. S. Corasaniti and J. Khoury, Phys. Rev. D 73, 083509 (2006).
\bibitem{30} L. Amendola, D. Tocchini-Valentini, Phys. Rev. D64,
043509 (2001); G. W. Anderson, S. M. Carroll,
ArXiv:astro-ph/9711288.
\bibitem{31} J. H. He, B. Wang, Y. P. Jing, JCAP 07, 030 (2009)
ArXiv:0902.0660.
\bibitem{z} G. Caldera-Cabral, R. Maartens, B. Schaefer,
arXiv:0905.0492.
\bibitem{50} J. Weller, A. M. Lewis, Mon. Not. Roy. Astron. Soc. 346, 987 (2003),
ArXiv:astro-ph/0307104.
\bibitem{51} R. Bean, O. Dore, Phys.Rev. D 69,
083503 (2004),ArXiv:astro-ph/0307100.
\bibitem{easy} M. Doran, JCAP 0510, 011 (2005).
\bibitem{BOOMERanG} C. J. MacTavish, et al., Astrophys. J.
  647, 799 (2006).
\bibitem{CBI} A. C. S. Readhead, et al., Astrophys. J.
  609, 498 (2004).
\bibitem{VSA} C. Dickinson, et al., Mon. Not. Roy. Astron. Soc.
353, 732 (2004).
\bibitem{ACBAR} C. L. Reichardt, et al., ArXiv:0801.1491.
\bibitem{SNeIa} A. G. Riess, et al., ArXiv:astro-ph/0611572.
\bibitem{SDSS} P. McDonald, et al., Astrophys. J. Suppl.
163, 80(2006); P. McDonald, et al., Astrophys. J. 635, 761 (2005).
\bibitem{field} S. Micheletti, E. Abdalla, B. Wang, Phys. Rev. D (in press)
  ArXiv:0902.0318.

%%%%%%%%%%%%%%%%%%%%%%%%%%%%%%%%%%5
\bibitem{ISWcross} Robert G. Crittenden, Neil Turok, Phys.Rev.Lett. 76
  (1996) 575; Uros Seljak, Matias Zaldarriaga, Phys.Rev. D60 (1999) 043504; A. Cooray,  Phys. Rev. D 65, 103510 (2002)

\bibitem{ISWmeasurements} S. Boughn, R. Crittenden, Nature 427 (2004)
  45-47; M.R. Nolta, et al., Astrophys.J. 608 (2004) 10-15; P.
  Fosalba, E.  Gaztanaga,
  Mon.Not.Roy.Astron.Soc. 350 (2004) L37-L41;
  P. Fosalba, et al.,
  Astrophys.J. 597 (2003) L89-92; R. Scranton, et al. 2003,
  astro-ph/0307335; P. Vielva, E. Martinez-Gonzalez, M. Tucci, astro-ph/0408252;
  N. Afshordi, Y.S. Loh, M. Strauss, Phys.Rev. D69 (2004)
  083524; Nikhil Padmanabhan, et al. Phys.Rev. D72 (2005) 043525; A. Cabre, et
  al., MNRAS 381, 1347 (2007); G. Olivares, et al.,
Phys. Rev. D 77 103520 (2008); B. M. Schaefer, MNRAS 388, 1403
(2008).
%%%%%%%%%%%%%%%%%%%%%%%%%%%%%%%%%%%%%%%
\bibitem{select} J. Lesgourgues, W. Valkenburg, E. Gaztanaga, Phys, Rev. D77 063505
(2008).
\bibitem{Giannantonio} T. Giannantonio, et al., Phys. Rev. D77, 123520
(2008); S. Ho, et al., Phys. Rev. D 78 043519 (2008).






\end{thebibliography}
\end{document}